\documentclass[manuscript]{aastex}

\begin{document}
\title{Observations of giant outbursts from Cygnus X-1}

\author{S. Golenetskii, R. Aptekar, D. Frederiks, E. Mazets, and V. Palshin}
\affil{Ioffe Physico-Technical Institute, St. Petersburg, 194021, Russia}
\email{golen@pop.ioffe.rssi.ru}

\author{K. Hurley}
\affil{University of California, Berkeley, Space Sciences Laboratory,
Berkeley, CA 94720-7450}

\author{T. Cline}
\affil{NASA Goddard Space Flight Center, Code 661, Greenbelt, MD 20771}

\author{B. Stern\altaffilmark{1}}
\affil{Institute for Nuclear Research, Moscow 117312, Russia}

\altaffiltext{1}{Stockholm Observatory, SE-133 36, Saltsj\H{o}baden, Sweden}

\begin{abstract}

We present interplanetary network localization, spectral, and time history information
for 7 episodes of exceptionally intense gamma-ray emission from Cygnus X-1.  The
outbursts occurred between 1995 and 2003, with durations up to $\thicksim$28000 seconds.  The observed
15 - 300 keV peak fluxes and fluences reached $\rm3 \times 10^{-7}\, erg \, cm^{-2} s^{-1} \, and 
\, 8 \times 10^{-4}\, erg \,cm^{-2}$
respectively.  By combining the triangulations of
these outbursts we derive an $\thicksim$ 1700 square arcminute (3 $\sigma$) error ellipse which contains Cygnus
X-1 and no other known high energy sources.  
The outbursts reported here occurred both when Cyg X-1 was in the hard state as well as in the soft one,
and at various orbital phases.  The
spectral data indicate that these outbursts display the same parameters as those of the underlying hard and soft states,
suggesting that they represent another manifestation of these states. 
 
\end{abstract}


\section{Introduction}

The X-ray source Cygnus X-1 was discovered by Bowyer et al (1965), and
its optical counterpart, a spectroscopic binary at a distance of
$\sim$ 2 kpc with a 5.6 day period (HDE226868),
was identified by Webster and Murdin (1972) and Bolton (1972).  The primary is
a supergiant, and the mass of the secondary is at least $\rm 7 M_{\sun}$ (Gies \& Bolton
1986), making it a black hole candidate and a high mass X-ray binary.  The X-ray source, which has been observed from
keV to MeV energies, is thought to be powered
by accretion (e.g. Petterson 1978).  Soft X-radiation may be produced in an accretion
disk close to the black hole, and hard X-rays by inverse Compton scattering in a separate hot
plasma; see Liang and Nolan (1984) for a review.  Recently, a relativistic
jet has been detected in the radio (Stirling et al. 2001), and Romero et al. (2002)
have suggested that Cygnus X-1 is a microblazar, that is, that we are observing it
close to the axis of the jet.

Cygnus X-1 is known to exhibit two states of X-ray emission, which are thought
to depend on the accretion rate (Esin et al. 1998).  The first,
more common one, is the hard state, in which the 20 - 200 keV flux is roughly one Crab, and
the $<$10 keV emission is about 0.5 Crab.  The second
is the soft state, in which the low energy X-ray flux increases by a factor of 2 - 4, while the high
energy flux decreases by a factor of about 2.  Transitions between the two states and flaring have been
documented extensively (Ling et al. 1997; Zhang et al. 1997; Cui et al. 2002; McConnell et al. 2002), and both states exhibit variability of a factor of two or so on all timescales.  

In this paper we report on seven episodes of long, intense gamma-ray emission from this source; we refer
to them as \it outbursts \rm (and identify them by their dates), 
to distinguish them from the shorter, less intense flaring which has been documented
previously.  One of these outbursts, 950325, was found by Mazets et al. (1995) and initially thought to
be a gamma-ray burst; subsequently, however, its source was found to be Cygnus X-1 by M. Briggs
(private communication, 1995).  Later, Stern et al. (2001) called attention to five outbursts 
on 1999 April 19-21.  During the two brightest events, the hard X-ray flux increased by over
an order of magnitude for $\sim$ 1000 s. Observations by \it Ulysses \rm
and Konus - \it Wind \rm of still three more outbursts in 2002 - 2003 have prompted us to
reanalyze all the available data in order to obtain better energy spectra and a more precise source location as well.  

\section{Observations}

The outbursts reported here were discovered in the time history
data of the gamma-ray burst (GRB) detectors aboard
the \it Wind, Ulysses, \rm and \it Compton Gamma-Ray Observatory \rm (CGRO) spacecraft (Konus, GRB, and the 
Burst and Transient Source Experiment, or BATSE,
experiments respectively).  In general, the
data used were for the non-triggered modes, with low temporal and spectral resolution (0.25 - 2.9 s and 3
channels, respectively).  Although in some
cases, the emission was intense enough to trigger BATSE, the data obtained in the triggered 
modes had too small a duration to cover the entire outburst, and the lower time resolution continuous data
have been used for most of our analysis.  Table 1 gives the dates, times, and durations of the seven events, as well as
the spacecraft which observed them, and the spectral state and orbital phase of Cyg X-1.  
To determine the source of the outbursts, we have employed the usual method
of interplanetary network (IPN) triangulation for a repeating source (e.g. Hurley et al.\, 1999).  In all cases, the
IPN consisted of only two widely separated spacecraft (\it Ulysses \rm and either CGRO or \it Wind \rm), 
so a single annulus of location was obtained for
each flare.  The seven annuli were combined statistically to produce the 1, 2, and 3 $\sigma$ error ellipses shown in figure 1.
The only known hard X-ray source in this region is Cyg X-1, which lies at the $\sim$ 95 \%  confidence level.

Although \it Ulysses \rm observed all seven of the outbursts in Table 1, it was in a mode in which
no energy spectra were accumulated.  Konus \it Wind \rm observed 6 of the 7 outbursts (one
outburst, 990421B, occurred during a period when the data were contaminated by solar
particles), and we have used the data from it exclusively for time-resolved spectral analysis.  
The time and energy resolutions vary depending on the mode of operation.  
For five outbursts we have data in three energy channels (nominally 10-50, 50-200, and 200-750 keV, or
G1, G2, and G3, respectively) which span the entire event.
In the sixth case (950325) a cosmic gamma-ray burst fortuitously triggered Konus during the outburst, allowing
it to record multi-channel energy spectra for a total of 360 s, starting 120 s after the GRB trigger,
when the 15 s long GRB had ceased emitting. In addition, for this event, there are also 3 channel data.
Because the most detailed data are available for it, we discuss this outburst first.

The 950325 outburst is presented in detail in figure 2.  In the top three panels, the time histories in the
G1, G2, and G3 windows are shown, and in the bottom panels, the hardness ratios G2/G1 and G3/G2.
The letters A-D indicate the intervals where energy spectra have been analyzed. (The results of
this and other analyses are summarized
in table 2.)  In particular, B marks the 360 s interval where multi-channel spectra were accumulated.  The
gap following B is the interval where the data were transferred to the onboard tape recorder. The hardness
ratios display a weak but obvious spectral variability.  From them, it is reasonable to assume that
the multi-channel energy spectra measured in interval B, during the fading stage of the flare, also characterize the spectrum
of the preceding, intense phase.  The multi-channel spectrum is presented in figure 3 in $\rm \nu F_\nu$ units.
(The spectral deconvolution procedure has been described by Terekhov et al. 1998).  The solid line
gives the best fit spectrum, $\rm E^2 A (E/1 keV)^{-\alpha}exp(-E/E_0)\,  keV \, cm^{-2} s^{-1}$,
with $\rm A=0.88 \pm 0.23 \,photons \, cm^{-2}\, s^{-1} \,keV^{-1}, \, the \, 
photon \, index \, \alpha = 1.24 \pm 0.15, \,and \, E_0=
112 \pm 30\, keV, and \, with \, \chi^2=22.6$ for 18 degrees of freedom.  This spectral shape is consistent with
that of the Cyg X-1 hard state.

The 15 - 300 keV time histories of the other outbursts are presented in figures 4 - 9.  The vertical
dotted lines and letters at the top of the Konus plots indicate the intervals over which 3 channel energy spectra
were accumulated to improve statistics.  Fits were done with the function 
$\rm  A (E/1 keV)^{-\alpha}exp(-E/E_0)  \,photons\, cm^{-2}\, s^{-1}\, keV^{-1}$, and the fitting
parameters are given in table 2.  E$_p$, the peak energy, was obtained from the spectrum
expressed in $\rm \nu F_\nu$ units.  The fluences and peak fluxes, given in table 3, were calculated
assuming no spectral evolution over the intervals considered.

It should be noted that an omnidirectional sensor detects an outburst from a source as a temporary
increase in the count rate over an average background level.  Clearly in this case that level includes
any persistent flux from Cyg X-1.  Thus the total flux during an outburst will be the sum of the
persistent and flaring fluxes detected by Konus.  However, the value of the persistent flux from
Cyg X-1 at time intervals adjacent to the outburst is poorly known.  Accordingly the fluences and peak
fluxes in table 3 are only to be considered estimates of the outburst component.

\section{Discussion}

Long-term observations of Cyg X-1 in X-rays and soft gamma-rays from HEAO 3 (Ling et al. 1987), BATSE (Ling et al. 1997), 
Mir-Kvant (Borkus et al. 1995), and BeppoSAX (Frontera et al. 2001) have shown that the source spends most of the time in the hard state. 
According to BATSE data for 1991 - 1999 (McConnell et al. 2002), the hard flux in the 20 - 100 keV  energy range varied between 0.2 and 0.36 photons cm$^{-2}$ s$^{-1}$, roughly the average value  corresponding to the $\gamma_2$ level in the 45 - 140 keV band identified by Ling et al. (1987). In 1999 April, several outbursts lasting up to 1000 s were 
found in the BATSE data (Stern et al. 2001). The peak fluxes in the 50 - 300 keV range, 0.3 - 1.1 photons cm$^{-2}$ s$^{-1}$, were significantly higher than the average value of  $\rm \sim 0.1 \,photons \,cm^{-2}\, s^{-1}$.  Transitions to the soft state lasting for several months were observed only a few times. According to the 
data of the All Sky Monitor aboard the \it Rossi X-ray Timing Explorer \rm
(RXTE/ASM, McConnell et al. 2002), the soft X-ray (2 - 10 keV) count rates increase from $\rm \sim 20 \,to \sim 80 \,counts \,s^{-1}$ after transitions to the soft state. From the BATSE and RXTE light curves we have ascertained that outbursts 950110, 950325, 990421A and B, and 030212 occured while Cyg X-1 was in the hard state whereas outbursts 020224 and 020331 arose from the soft state. 

Broadband spectral observations of Cyg X-1 from numerous spacecraft and combinations of spacecraft have been used to study comprehensively
the spectral shapes at various times in order to deduce the emission mechanism and the geometry of the regions emitting and reprocessing
radiation (Mir-Kvant: Borkus et al. 1995; CGRO: Ling et al. 1997; Ginga and CGRO: Gierlinski et al. 1997; RXTE: Dove et al. 1998;
ASCA, RXTE, and CGRO: Gierlinski et al. 1999; BeppoSAX: Frontera et al. 2001; and CGRO and BeppoSAX: McConnell et al. 2002).
The accuracy of our data is insufficient for such a thorough analysis of source models.  Rather, we have focused on obtaining the essential information on a new feature of Cyg X-1 activity from limited observational data using the simplest model spectra, namely a power law, 
$\rm AE^{- \alpha}$, or a power law with an exponential cutoff,  $\rm AE^{- \alpha}exp(-E/E_0)$. 
Apart from the lack of high resolution spectral data, the most serious difficulties in our data processing are connected with slow changes in 
the background level, especially in the low energy ranges. Background count rates in the G1 window and, to a lesser extent,
in G2, are strongly affected by the activity of cosmic X-ray sources and vary, even when the Sun is quiet, by $\pm1 \%$,
averaged over hours and days. For this reason, for example, we do not attempt spectral fits for the long, weak onsets of outbursts 950110 and 950325.

The intense stage of each outburst was subdivided into several time
intervals
in some concordance with the temporal behaviour of the hardness ratio,
as shown in figures 2 and 4-9. Spectral parameters for all of these intervals
were determined using the following procedure.
The counts accumulated in energy band G$_i$ (i=1, 2, 3) can be represented
as an integral over an energy loss spectrum
$$ N_{G_i} = \int_{G_i} \left[ \mathbf{M \cdot f} (A, E_0, \alpha)
\right] (E) \mathrm{d} E $$
using a model photon spectral function $\mathbf{f}$ and the detector
response matrix $\mathbf{M}$. The resulting system of three equations with
unknowns
A, $\alpha$, and E$_0$ was solved by the Marquardt-Levenberg method
and the uncertainties in the parameters were calculated from the covariance matrix.
In the case of the exponentially attenuated power law,
parameters $\alpha$ and E$_0$ are strongly correlated. The peak energy
E$_p$
for a $\nu F_\nu$ spectrum is a more robust characteristic.
Estimates of the errors in the spectral parameters were checked by numerical
simulations,
i.e. by introducing Poisonnian-distributed deviations to the
count rates G1, G2, and G3,
applying the procedure described above,
and analyzing the distributions of the spectral parameters. 
The results of the simulations are in very good agreement with estimates 
from the covariance matrix.

All of these data are collected in table 2, which includes an additional interval B* for 950325.
This interval,  where G1, G2, and G3 data were available, partially overlaps interval B. 
From this it is evident that the parameters obtained by both methods agree well to within their uncertainties. 
The 15 - 300 keV fluences and peak fluxes for all the outbursts are presented in table 3. 

The main conclusions which can be drawn from these data are the following. 

\begin{itemize}

\item An exponential cutoff is essential when fitting outbursts arising
from the hard state.
The parameters $\alpha$ and E$_0$ for these outbursts
are consistent with the values for the hard state of Cyg X-1
reported in literature.

\item The spectra of the 2002 soft state outbursts
are softer in our energy range.
The photon spectra of 020331 are well fit
by a single slope power law with $\alpha \sim 2.6$, which
is certainly softer than any spectrum observed in the hard state.
The spectra of 020224 have an intermediate shape
and display a low E$_0$ with large uncertainties in $\alpha$.

\item  The hardness ratio G2/G1 varies slightly during outbursts, but it does not display
any clear correlation with the luminosity.  The only
exception is observed in the outburst of 990421B, where the hardness decreases
substantially at the peak (see Stern, Beloborodov, \& Poutanen 2001)

\item  In general, it seems that giant outbursts maintain the spectral parameters, and hence emission mechanism,
of the current underlying spectral state of Cyg X-1,  
even though the 15 - 300 keV peak fluxes 
are up to ten times higher than the persistent emission fluxes.  The corresponding fraction of the peak luminosity 
in these outbursts approaches one tenth of the Eddington luminosity for an object with $\rm M/M_\sun \sim 10$.

\end{itemize}

The soft and hard states of Cygnus X-1 and other galactic black hole candidates have comparable
bolometric luminosities, but different energy spectra (e.g. Ling et al. 1997; Zdziarski et al. 2002).  
The hard X-ray spectra of these sources are typically characterized
by power law with exponential cutoff with photon index $\alpha \sim
-1.6$
and cutoff energy $E_0 \sim 150$ keV (Phlips et al. 1996, Gierlinski et
al. 1997,
Dove et al. 1998), which are consistent with those in table 2
for the first four outbursts and the seventh one.  Many authors have interpreted them as
thermal Comptonization  
with optical depth $\tau_T \sim 1$ and electron temperature 
$\sim 50 - 100 $ keV (for recent studies see Maccarone \& Coppi, 2002). 

The model for transitions between the soft and hard states discussed in a number
of recent works is based on the idea of 
accretion disk transformations. In the soft state the accretion proceeds 
via a standard optically thick accretion disk (Shakura \& Sunyaev, 1973)
emitting a blackbody soft X-ray peak and a nonthermal tail resulting from
coronal activity in the disk.  In the 
hard state the inner part of the disk is thought to form a
hot geometrically thick corona (see e.g. Barrio, Done \& Nayakshin, 2002, 
Poutanen, Krolik \& Ryde, 1997, 
and references therein) with a moderate ($\tau_T \sim 1$) optical depth. 
The outer region of the disk, beyond a few tens to one hundred gravitational 
radii, still remains optically thick, emitting a cooler (0.1 - 0.2 keV) and 
weaker blackbody component. The physical grounds for
such a transformation could be a transition of the inner part of the disk
into a regime balancing between ADAF and standard disk modes 
(Esin et al. 1998). However such a regime 
requires a fine tuning which would clearly be violated during giant 
outbursts (Stern, Beloborodov \& Poutanen, 2001).

An alternative to disk accretion for the case of the hard 
state could be accretion of the companion wind with a near-critical 
angular momentum (Illarionov \& Sunyaev, 1975). Then,  instead of a standard 
optically thick disk, a small-scale disk with $\tau_T \sim 1$ can appear.
This scenario can naturally describe the hard state 
spectrum (Beloborodov \& Illarionov, 2001)
in terms of thermal Comptonization in the small disk within
a few gravitational radii. No difficulties with
the large observed variabilities arise in this model. 

The absence of a clear correlation between luminosity and hardness is
noteworthy.  Actually, the presence of a luminosity-hardness correlation
or anti-correlation depends on many details, including the geometry of
the emission region, the soft radiation, and pair production.  If the
Comptonizing plasma is pair-dominated, then the temperature is a weak
function of the luminosity: pair production provides a thermostatic
effect (see, e.g. Malzac, Beloborodov, \& Poutanen, 2001).  The spectrum
in this case is softer at higher luminosities (if the geometry is fixed),
but the dependence is weak (Stern et al. 1995).  A luminosity-hardness
anti-correlation is expected, but we do not observe it because of
statistics. 

What triggers the giant outbursts? The normal time variability of Cygnus X-1
has a broad power density spectrum which probably arises from instabilities
in the disk.  The outbursts do not appear to display the same variability
pattern; moreover, they show similar temporal behaviors in the hard and
soft states, which correspond to different states of the disk.
Therefore, they are unlikely to originate due to an intrinsic
disk effect. The most straightforward suggestion, and the one which
we adopt as a working hypothesis,
is that this could be some rare eruptive phenomenon in the donor wind ejection.
The typical dynamical timescale for standard disk accretion in the 
case of Cyg X-1
is much longer than the duration of the outbursts, so that any fluctuations 
in the donor ejection rate would be dilated by up to a day or more (Bisnovaty-Kogan,
private communication, 2002). Direct wind accretion, on the other hand,   
has a dynamical timescale $\sim 1000$ s, but a shorter timescale can arise from 
instabilities in the accreting flow (Illarionov, private communication, 2002).

Other interpretations are possible. For example, the outbursts could be attributed to the recently
discovered jet of Cyg X-1 (Stirling et al. 2001) and their recurrence 
could be explained by its precession (Romero et al. 2002).

\section{Conclusion}

The outbursts reported here have durations comparable to the period of a low Earth-orbiting spacecraft, making them difficult
to detect and follow from experiments aboard such spacecraft, but relatively easy for experiments which are
far from Earth and do not undergo occultation and orbital background variations.  Indeed, our observations of 990421A
show that it continues well beyond the point at which it was Earth-occulted to BATSE, and our observations of
990421B indicate that it commenced several hundred seconds before it rose on BATSE (Stern et al. 2001).  These observations
point to a new use for the IPN, namely tracking long, intense flares from Galactic transients.  The data from a single
experiment which has little or no directional and/or spectral information are easily confused with solar X-ray and particle events,
which explains why it has taken so long to determine the origin of some of the outbursts presented here.  However, confirmation
by a second spacecraft solves this problem and gives an annulus of position which in many cases may be sufficient to determine
the origin of the emission.   For several outbursts, the BATSE
observations alone yield source directions which are only accurate to 17$\degr$ ,  leaving the possibility that the source could have been
Cygnus X-3.  However, relatively small error ellipses may be obtained from multiple observations, and our localizations rule this possibility out conclusively.

The histories of Cygnus X-1 and GRBs have been curiously intertwined over the decades.  In the
early days of X-ray astronomy, 
the use of interplanetary spacecraft was suggested to localize sources with rapid time variations such as Cyg X-1 (Giacconi 1972).  Although,
to our knowledge, such measurements were never carried out on persistent X-ray sources, today the technique is the basis of
the IPN, and the present observations demonstrate the feasibility of this suggestion.  Later, Mason et al. (1997) pointed out that bursts from Cyg X-1 could 
appear in the BATSE database; BATSE has indeed triggered many times on this source.  Some of
the long outbursts presented here
have time histories which, if compressed in time,
would be virtually impossible to distinguish from those of GRBs, and their spectra are hard.  These similarities suggest
that accretion onto a black hole, which is believed to power both Cyg X-1 and GRBs albeit under very different
circumstances, may manifest itself in similar ways in very different settings.

\section{Acknowledgments}

Support for the \it Ulysses \rm GRB experiment is provided by JPL Contract 958056,
and for Konus-\it Wind \rm by a Russian Aviation and Space Agency contract
and RFBR grant N 03-02-17517.  We are grateful to 
Al Levine and the ASM/RXTE team for quick-look results on Cygnus X-1, and to
Andrei Illarionov for useful discussions.  This research has made use of the SIMBAD database,
operated at CDS, Strasbourg, France.

\clearpage

\begin{figure}
\figurenum{1}
\epsscale{0.8}
\plotone{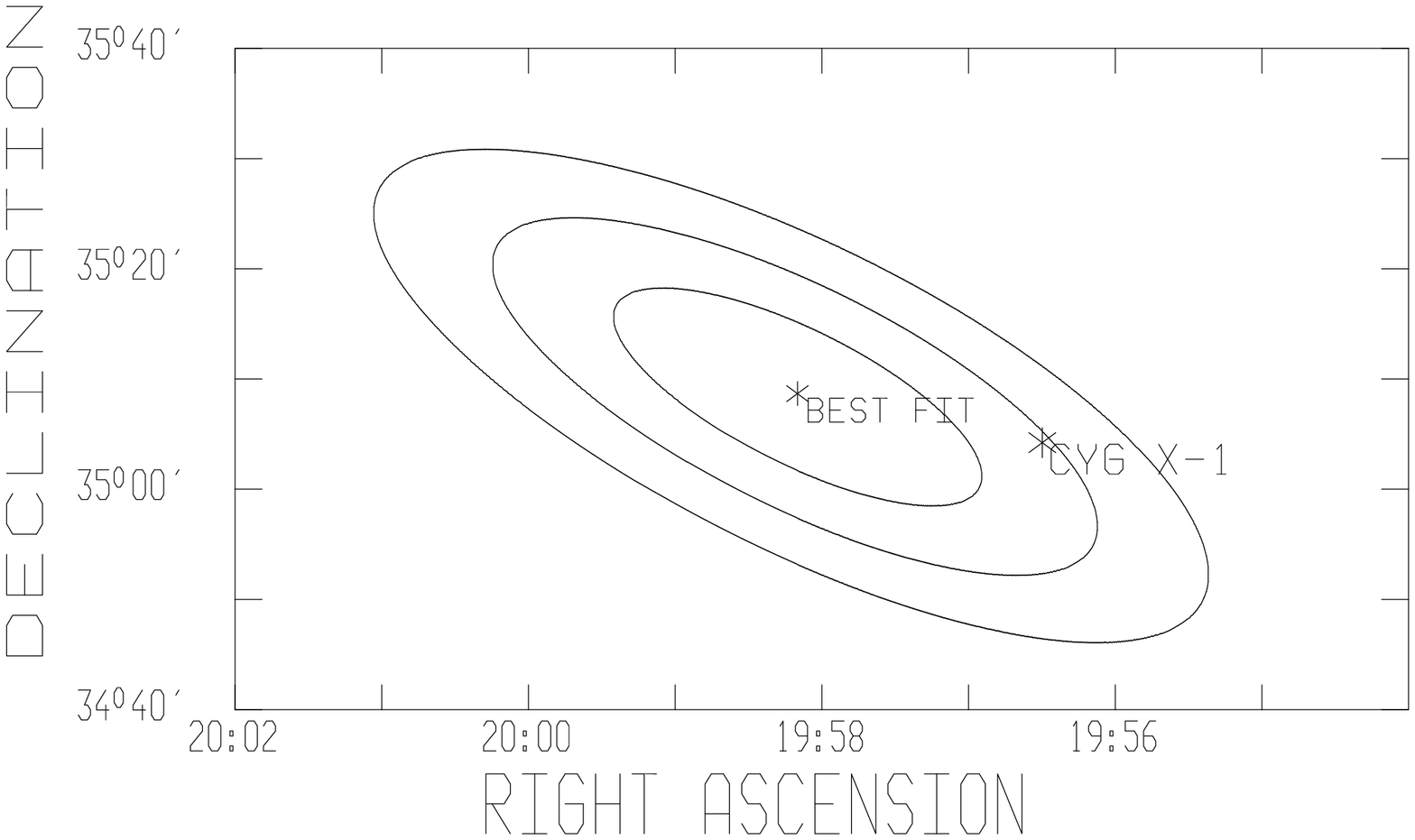}
\caption{1, 2, and 3 $\sigma$ error ellipses for the seven Cyg X-1 outbursts.  Their areas are 328, 887, and 1700
square arcminutes.  The best fit position is at $\rm \alpha(2000)=19^h 58^m 09^s, \delta(2000)=
35^o 08\arcmin 20\arcsec$, with a $\chi^2$ of 7.24 for 5 degrees of freedom
(7 annuli minus two fitting parameters).  Cyg X-1 lies 0.35$\degr$ away, at the $\sim$95 \% confidence level.
A search through the SIMBAD database has not identified any other X-ray sources within the
3$\sigma$ error ellipse.
}
\end{figure}

\clearpage

\begin{figure}
\figurenum{2}
\epsscale{1.}
\vspace{-2.in}
\plotone{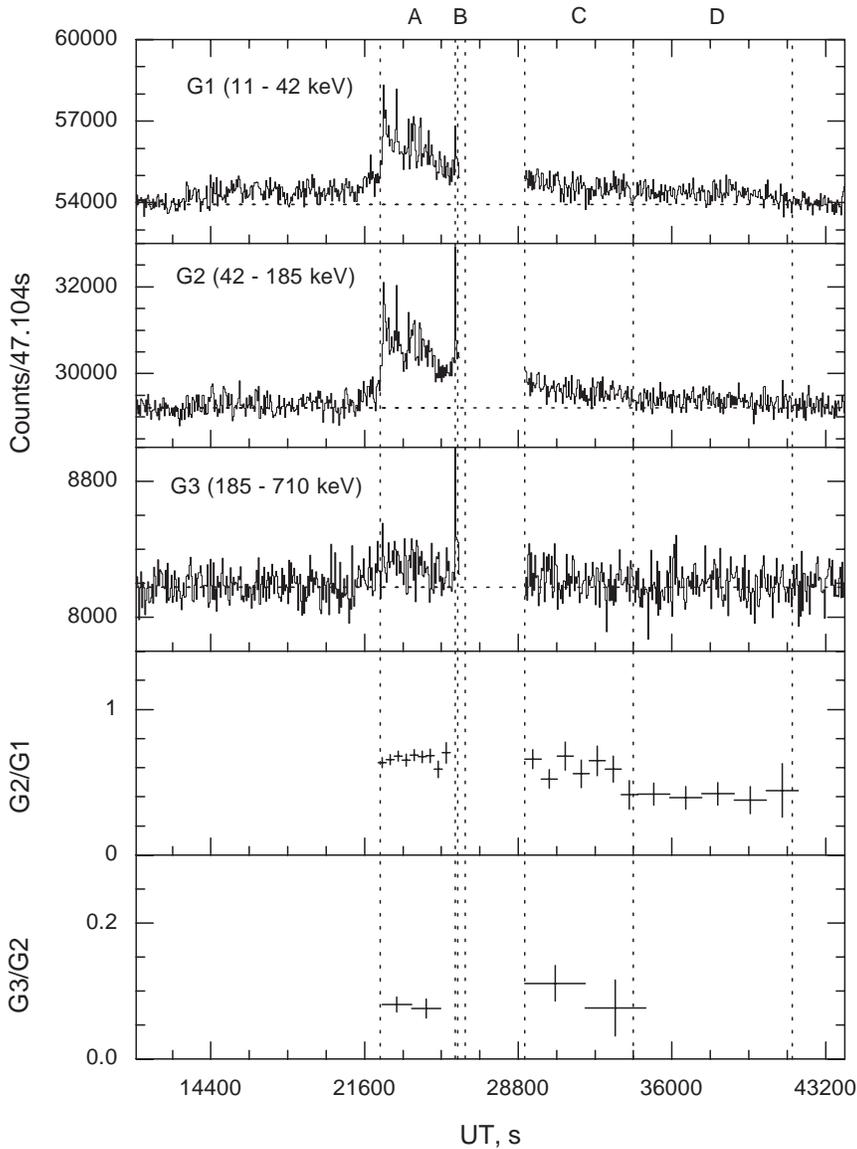}
\vspace{-2.5in}
\caption{The Cyg X-1 outburst on 950325 as observed by Konus-\it Wind \rm. The time histories are shown in 
the three energy bands G1, G2, and G3 with time resolution 47.104 s; the hardness ratios G2/G1 and G3/G2
are also shown. 
A, B, C, and D denote the time intervals where the spectral parameters were estimated. The backgrounds in G1-G3 
were slightly unstable due to an apparent Forbush decrease
which began on March 23 and lasted through March 25, according to the Konus data.  Thus the apparent weak onset
of the event prior to interval A cannot be definitely attributed to Cyg X-1.
The narrow spike at 25841 s in the beginning of interval B is the GRB which triggered a series of multichannel 
spectral measurements, covering the entire interval B. The gap after B was caused by the
transmission of high resolution data to the onboard tape recorder.  Background levels are indicated by the
dashed lines.}
\end{figure}

\clearpage

\begin{figure}
\figurenum{3}
\epsscale{1.3}
\vspace{-2.5in}
\hspace{-2.in}
\plotone{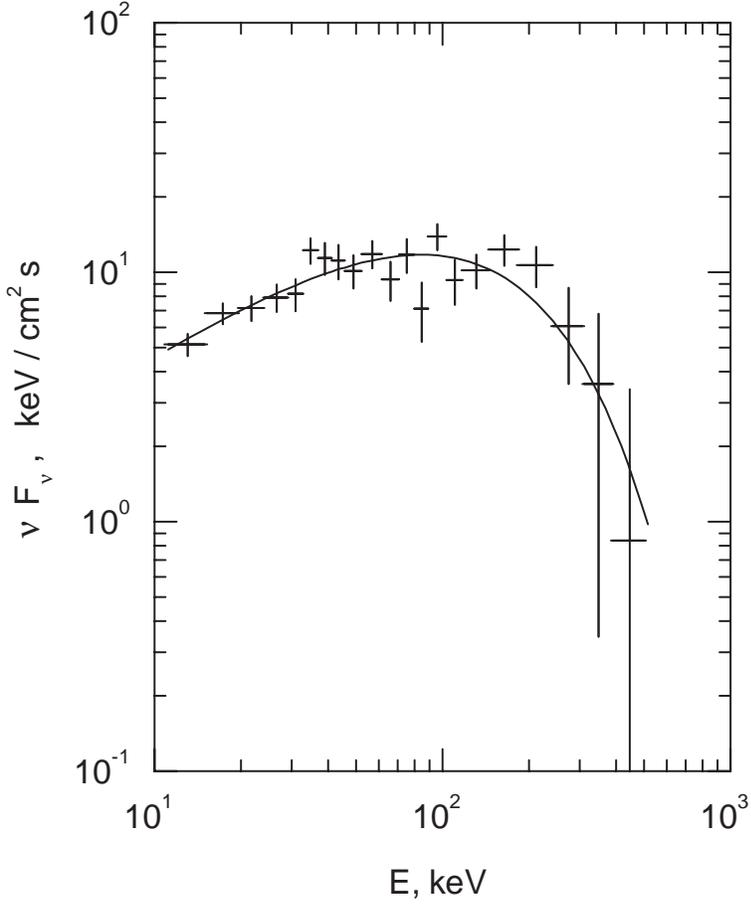}
\vspace{-3.5in}
\caption{Konus $\rm \nu F_{\nu}$ spectrum of the outburst on 950325. The spectrum was accumulated during the decay stage 
of the flare over 360 s of interval B with moderate statistics. The solid curve is the result of the spectral fit; $\rm \chi^2=$ 22.6 
for 18 degrees of freedom. The spectral shape is consistent with that of the hard state of Cyg X-1.}
\end{figure}

\clearpage

\begin{figure}
\figurenum{4}
\epsscale{1.2}
\vspace{-1in}
\plotone{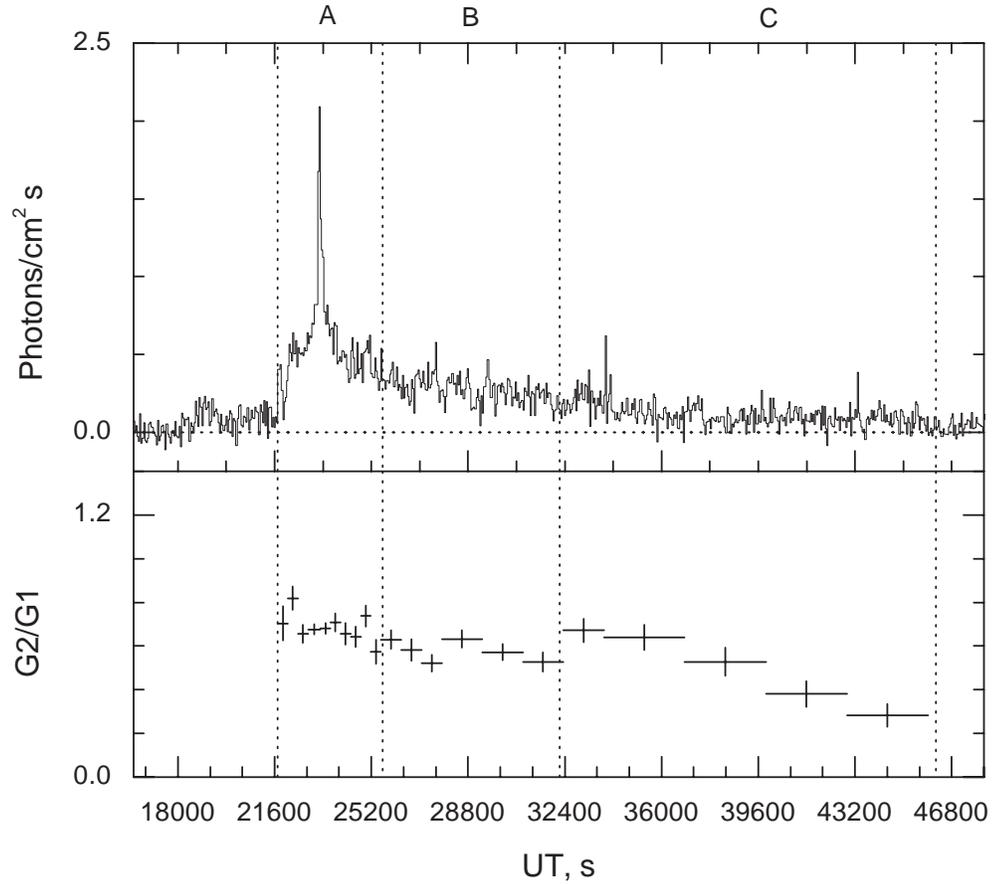}
\vspace{-3.5in}
\caption{Konus 15 - 300 keV photon fluxes (background subtracted),
and hardness ratios for the outburst on 950110.
Spectral fits were done for intervals A, B, C, and A+B+C. 
The photon fluxes in photons cm$^{-2}$ s$^{-1}$ were calculated by
deconvolving the count rate
data. The time resolution is 47.104 s.}
\end{figure}

\clearpage

\begin{figure}
\figurenum{5}
\epsscale{1.2}
\vspace{-1in}
\plotone{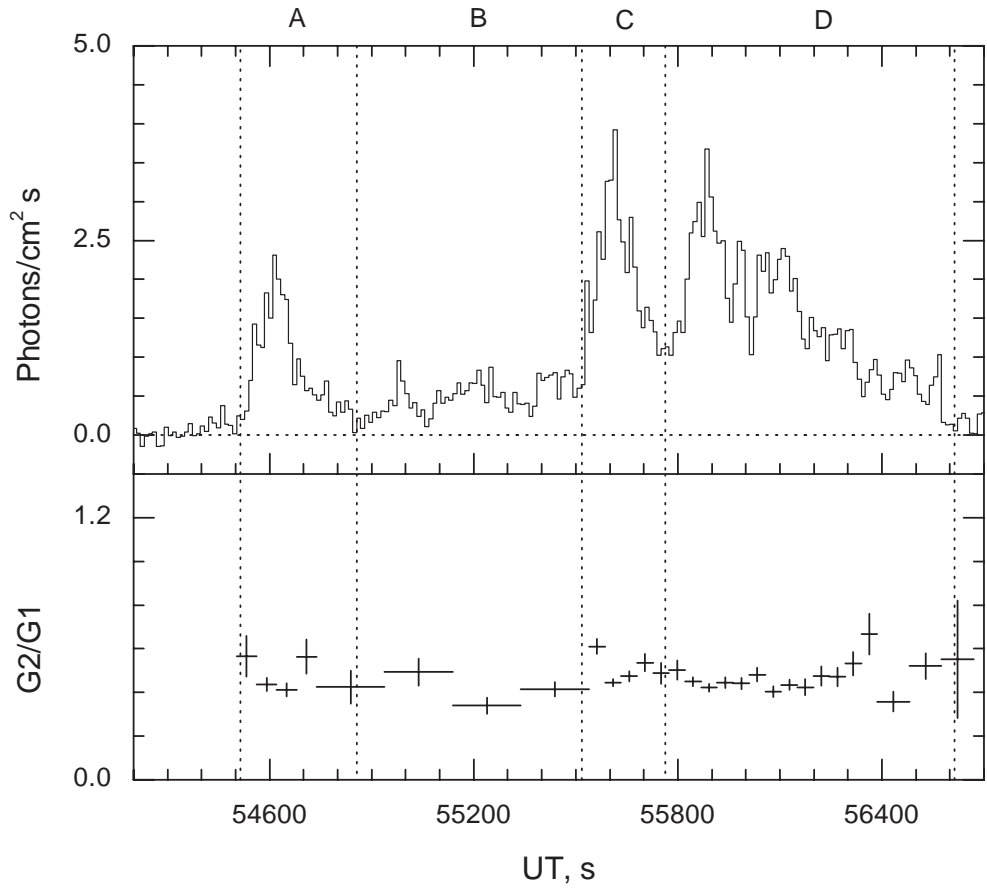}
\vspace{-3.5in}
\caption{Konus 15 - 300 keV photon fluxes (background subtracted)
and hardness ratios for the outburst on 990421A. The
time resolution is 11.776 s.  Spectral fits were done for
intervals A, B, C, D, and A+B+C+D.}
\end{figure}

\clearpage
\begin{figure}
\figurenum{6}
\epsscale{1.2}
\plotone{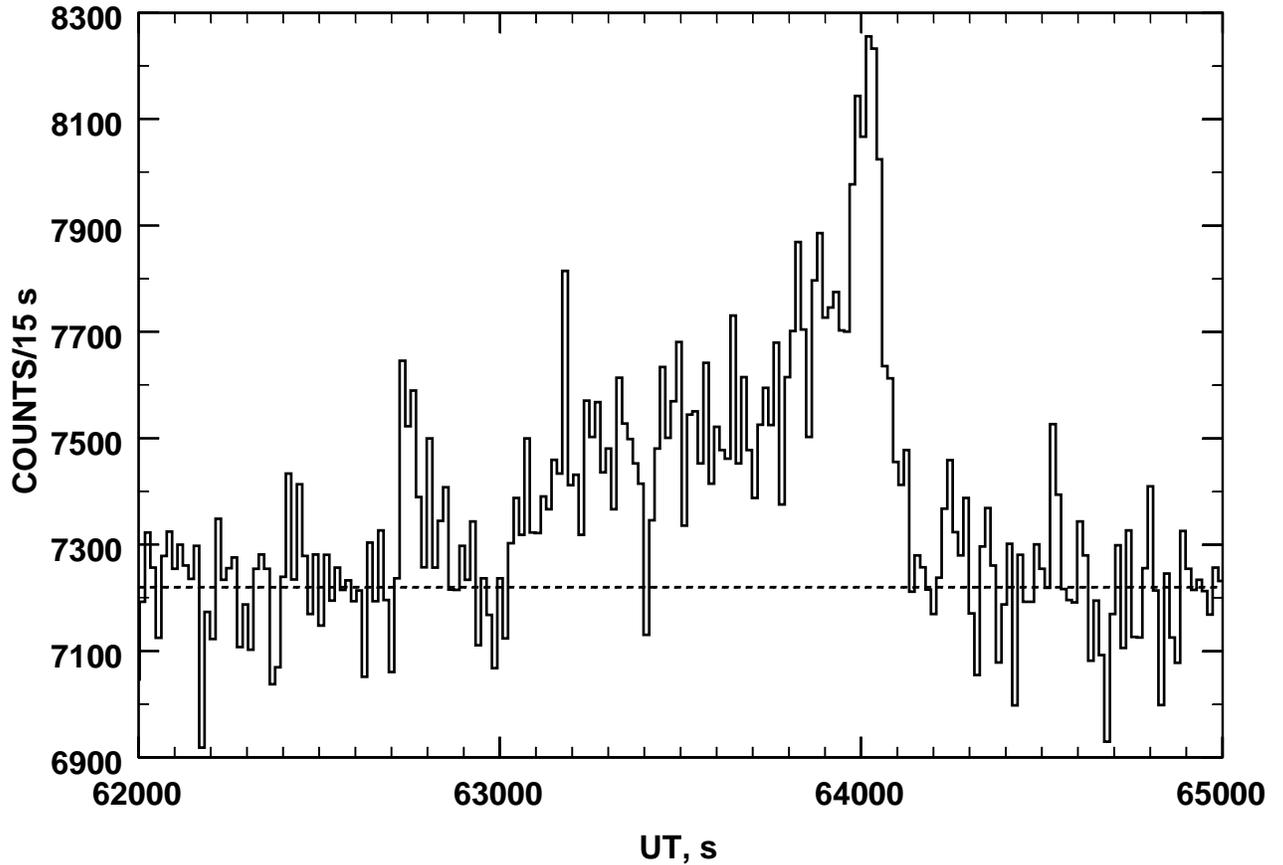}
\caption{\it Ulysses \rm 25 - 150 keV count rates for the outburst on 990421B. The time 
resolution is 15 s.  Note that the onset is several hundred seconds prior to the onset
at BATSE (Stern et al. 2001), due to Earth occultation.  No \it Ulysses \rm spectral data are available
for this outburst.  Due to the proximity in time to the previous outburst, the two could be considered
manifestations of a single episode, although the Konus and \it Ulysses \rm count rates remained
at their background levels between them.   In contrast to the outbursts on 950110, 950325, 990421A,
and 020224, this time history exhibits a slow rise and a fast decay.
The background level is indicated by the dashed line.}
\end{figure}

\clearpage

\begin{figure}
\figurenum{7}
\epsscale{1.2}
\vspace{-0.5in}
\plotone{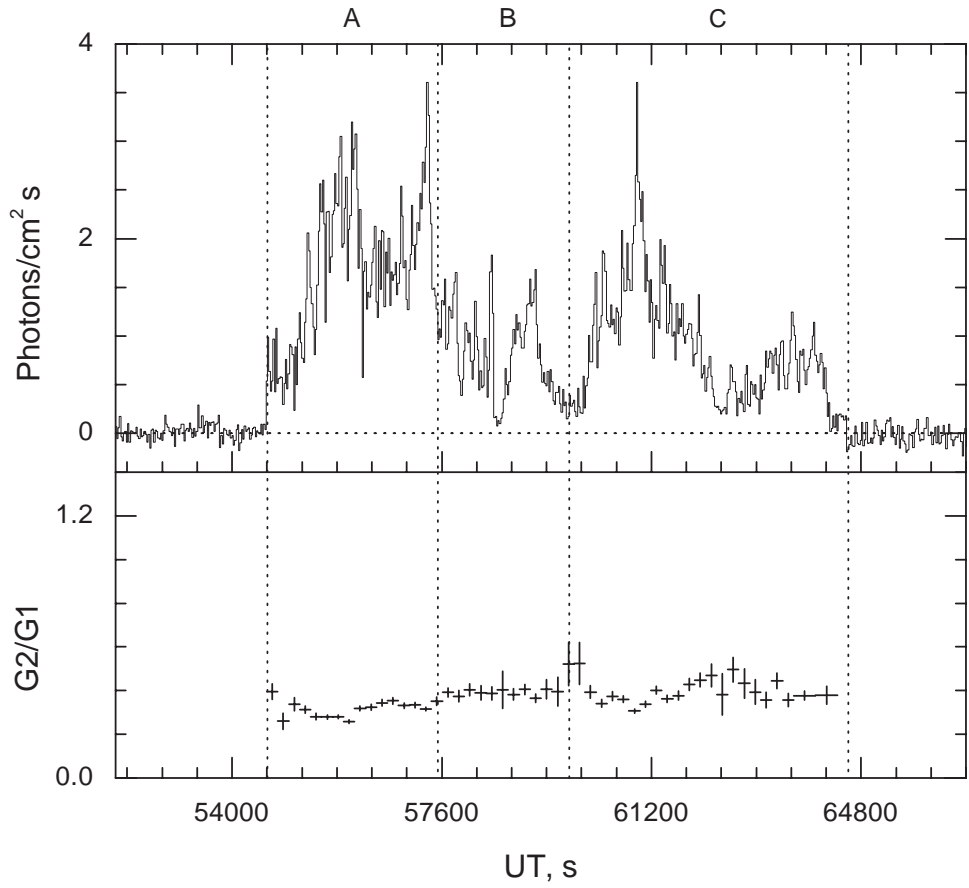}
\vspace{-4.5in}
\caption{Konus 15-300 keV photon fluxes (background subtracted)
and hardness ratios for the outburst on 020224.  The time
resolution is 23.552 s.  Spectral fits were done
for intervals A, B, C, and A+B+C.}
\end{figure} 

\begin{figure}
\figurenum{8}
\epsscale{1.2}
\vspace{-0.5in}
\plotone{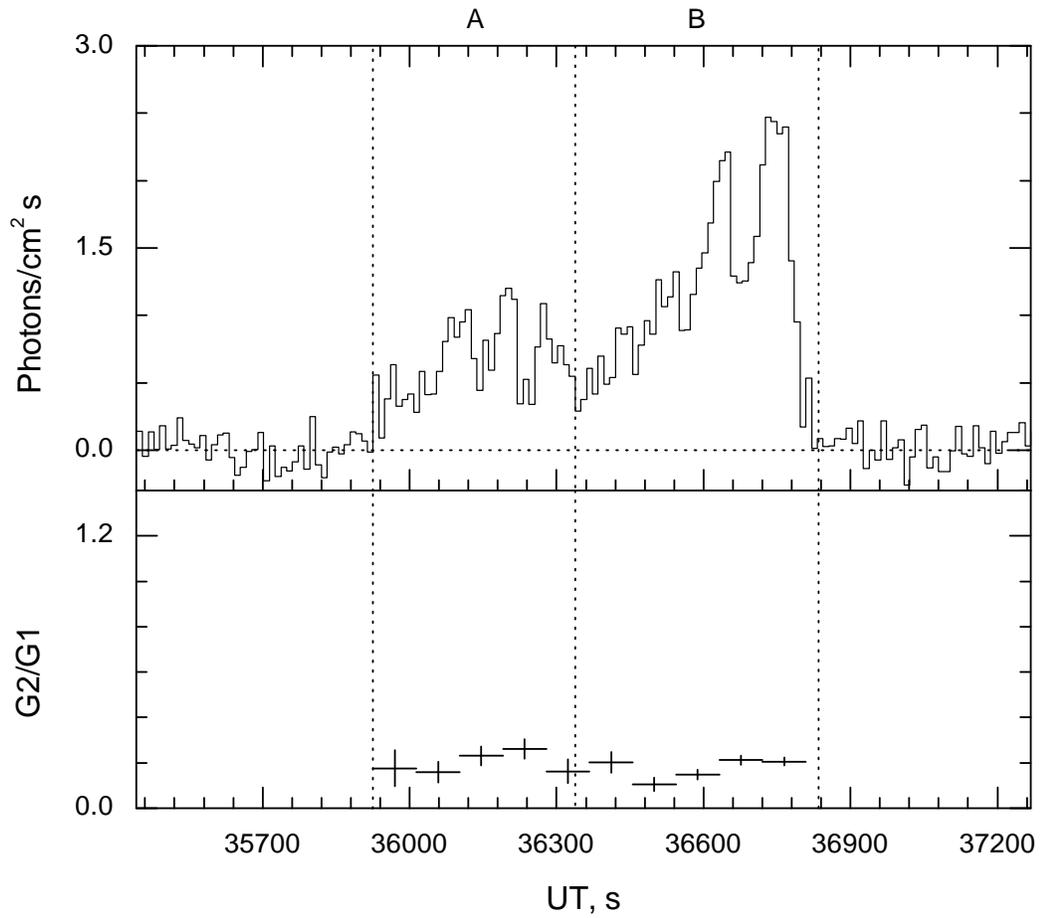}
\vspace{-4.5in} 
\caption{Konus 15-300 keV photon fluxes (background subtracted)
and hardness ratios for the outburst on 020331.
The time resolution is 11.776 s.  
Spectral fits were done for intervals A, B, and A+B. 
Note the resemblance to the time history of 990421B (slow rise, fast
decay).
}
\end{figure}
 
\clearpage

\begin{figure}
\figurenum{9}
\epsscale{1.2}
\vspace{-0.5in}
\plotone{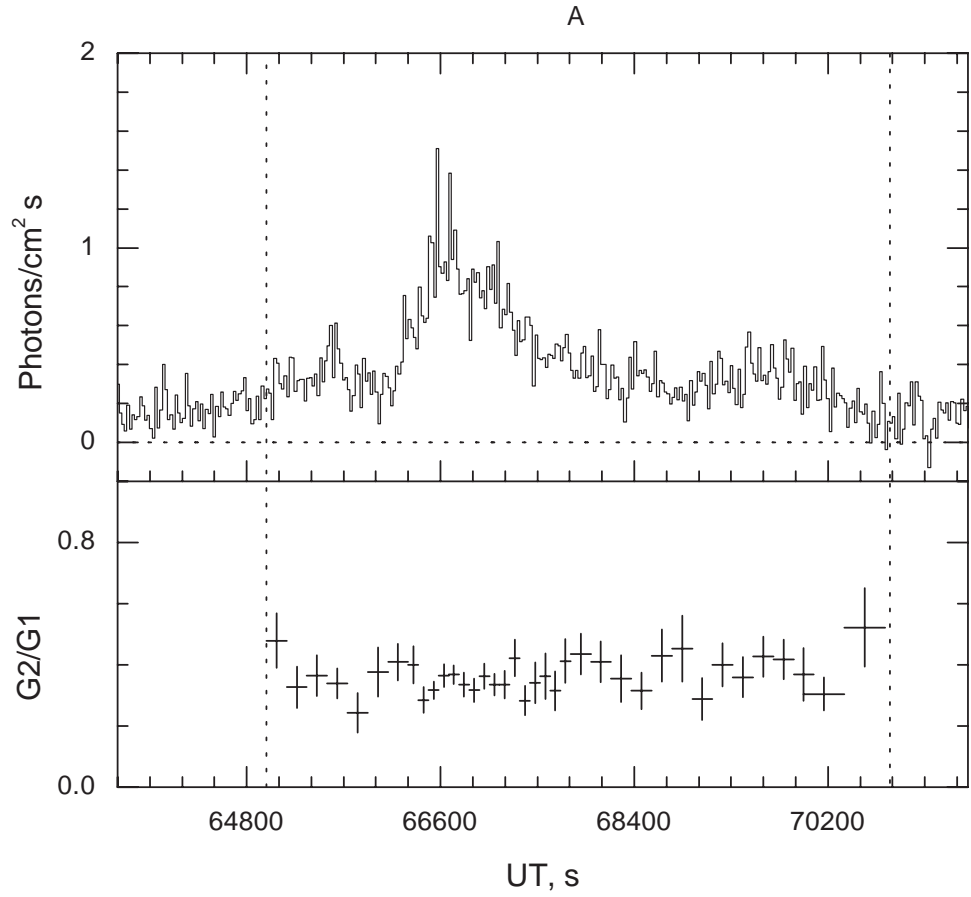}
\vspace{-4.5in} 
\caption{Konus 15-300 keV photon fluxes (background subtracted)
and hardness ratios for the outburst on 030212.  The time
resolution is 47.104 s.  The spectral fit was done
for interval A.}
\end{figure}

\clearpage

\begin{deluxetable}{lccccc}
\tabletypesize{\small}
\tablewidth{0pt}
\tablecaption{\it Seven Cygnus X-1 outbursts}
\tablehead{
\colhead{Date} &\colhead{Onset UT} &\colhead{Duration} 
&\colhead{Spacecraft } &
\colhead{Underlying} & \colhead{Orbital phase\tablenotemark{f}} \\
& \colhead{(approx.)} & \colhead{(approx.)} & & \colhead{spectral state}
& \colhead{(at onset)}\\
\colhead{YYMMDD}& \colhead{s} & \colhead{s} & & &
}
\startdata

950110 & 18600 & 27500 & \it Ulysses \tablenotemark{a}\rm\ , Konus
\tablenotemark{a,b}\ \ \ ,
BATSE \tablenotemark{c} & hard\tablenotemark{g} & 0.37 \\

950325 & 14400 & 28000 &  \it Ulysses \tablenotemark{a}\rm\ , Konus
\tablenotemark{a,b}\ \ \
, BATSE \tablenotemark{c} & hard\tablenotemark{g} & 0.57 \\
 
990421A & 54510  & 4870 &  \it Ulysses \rm, Konus, BATSE
\tablenotemark{d} & hard\tablenotemark{h} & 0.38 \\

990421B\tablenotemark{e} & 63100  & 1360 &  \it Ulysses \rm, BATSE
\tablenotemark{d} & hard\tablenotemark{h} & 0.40 \\

020224 & 54600 & 10000 & \it Ulysses \tablenotemark{a}\rm\ , Konus
\tablenotemark{b}  & soft\tablenotemark{h} & 0.10 \\

020331 & 35925 & 910 & \it Ulysses \rm, Konus & soft\tablenotemark{h} &
0.31 \\

030212    & 65000 & 5800     & \it Ulysses \rm, Konus &
hard\tablenotemark{h} & 0.16 

\enddata

\tablenotetext{a} {Golenetskii et al. (2002a,b)}
\tablenotetext{b} {Mazets et al. (1995)}
\tablenotetext{c} {Schmidt (2002)}
\tablenotetext{d} {Stern, Beloborodov, and Poutanen (2001)}
\tablenotetext{e} {This event may be a continuation of the preceding
one.}
\tablenotetext{f} {The phases were calculated using the orbital
ephemeris from Brocksopp et al. (1999).
Zero phase corresponds to the time of superior conjunction of the black
hole.}
\tablenotetext{g} {The state according to BATSE data (McConnell
et al. 2002).}
\tablenotetext{h} {The state was derived using the online data of RXTE
(http://xte.mit.edu/ASM\_lc.html).}

\end{deluxetable}

\clearpage

\begin{deluxetable}{cccccc}
\tabletypesize{\footnotesize}
\tablecaption{\it Energy spectra from Konus - Wind}
\tablewidth{0pt}
\tablehead{
\colhead{Date} & \colhead{Time interval}      &\colhead{A,}     &\colhead{$\alpha$}
&\colhead{E$_0$, keV} &  \colhead{E$_p$, keV}  \\
\colhead{}        & \colhead{(see figures 2, 4, 5, 7-9)} &\colhead{photons cm$^{-2}$ s$^{-1}$ keV$^{-1}$} &\colhead{}
& \colhead{}   & \colhead{}                   \\
}
\startdata

950110  &A                              &0.92$\pm$0.08    &1.24$\pm$0.05 &111$\pm$10                 &85$\pm$3    \\
        &B                              &0.93$\pm$0.12    &1.45$\pm$0.07 &133$\pm$22             &73$\pm$4      \\
        &C                              &1.07$\pm$0.15    &1.81$\pm$0.08 &452$\pm$218             &84$\pm$17      \\
        &A+B+C                  &0.83$\pm$0.06    &1.50$\pm$0.04        &162$\pm$16 &82$\pm$3      \\
950325  &A                              &0.97$\pm$0.09    &1.26$\pm$0.05 &112$\pm$11             &82$\pm$3               \\
        &B\tablenotemark{a}             &0.88$\pm$0.23    &1.24$\pm$0.15 &112$\pm$30             &85$\pm$9      \\
        &B\tablenotemark{b}                 &1.25$\pm$0.30        &1.36$\pm$0.14 &137$\pm$41             & 87$\pm$11    \\
        &C                              &0.61$\pm$0.12    &1.48$\pm$0.11 &156$\pm$44             &81$\pm$9      \\
        &D                              &0.94$\pm$0.47    &1.80$\pm$0.26 &163$\pm$128    &33$\pm$19      \\
        &A+C+D                  &0.67$\pm$0.07    &1.44$\pm$0.06        &133$\pm$17 &75$\pm$4      \\
990421A &A                              &1.35$\pm$0.92      &1.20$\pm$0.39        & 87$\pm$48       &69$\pm$5      \\
        &B                              &4.83$\pm$0.94     &1.83$\pm$0.12 &434$\pm$281    &74$\pm$20      \\
        &C                              &9.88$\pm$0.74    &1.63$\pm$0.05 &303$\pm$54         & 113$\pm$8        \\
        &D                              &3.08$\pm$0.53     &1.33$\pm$0.10 &111$\pm$19     &75$\pm$3      \\
        &A+B+C+D                        &4.24$\pm$0.41     &1.52$\pm$0.06 &156$\pm$20     &72$\pm$2      \\      
020224  &A                              &2.28$\pm$1.29     &1.10$\pm$0.34        & 63$\pm$21       &57$\pm$3      \\
        &B                              &0.48$\pm$0.40     &0.89$\pm$0.52        & 64$\pm$32       &71$\pm$2      \\      
        &C                              &2.73$\pm$0.47     &1.44$\pm$0.11 &121$\pm$23     &68$\pm$2      \\
        &A+B+C                  &2.38$\pm$0.44     &1.29$\pm$0.11        &88$\pm$13 &62$\pm$1      \\
020331  &A                              &57$\pm$5     &2.50$\pm$0.08   & &               \\
        &B                              &155$\pm$9      &2.60$\pm$0.05   & &               \\
        &A+B                            &107$\pm$5     &2.55$\pm$0.04        & &               \\
030212  &A   &0.93$\pm$0.47    &1.31$\pm$0.30 &90$\pm$37                 &61$\pm$2    
\tablenotetext{a}{Multichannel spectrum}
\tablenotetext{b}{3 channel spectrum}
\enddata
\end{deluxetable}

\clearpage

\begin{deluxetable}{lccccc}
\tabletypesize{\footnotesize}
\tablewidth{0pt}
\tablecaption{\it Fluences and peak fluxes from Konus - \it Wind \rm}
\tablewidth{0pt}
\tablehead{
\colhead{Date} & \colhead{Time interval}  & \colhead{15 - 300 keV
fluence}  &
\multicolumn{2}{c} {15 - 300 keV peak flux} & \colhead{Peak luminosity
at 2 kpc}            \\
                         & \colhead{s UT} & \colhead{$\rm erg \,
cm^{-2}$}  & \colhead{$\rm photons \, cm^{-2} \, s^{-1}$}  &
\colhead{$\rm erg \,
cm^{-2}\,s^{-1}$} &  \colhead{$\rm erg \, s^{-1}$} \\
}
\startdata
950110 & 21712 - 46215  & $\rm 4.5\times 10^{-4}$ & 2.4  & $\rm
1.9\times 10^{-7}$ &
$\rm 9.1\times 10^{37}$ \\
950325 & 22331 - 42635  & $\rm 2.8\times10^{-4}$ &  1.3  & $\rm
9.5\times 10^{-8}$ & $\rm
4.6\times 10^{37}$ \\    
990421A & 54514 - 56613  & $\rm 1.8\times10^{-4}$ &  3.8  & $\rm
3.0\times 10^{-7}$ & $\rm
1.4\times 10^{38}$ \\       
020224 & 54609 - 64583  & $\rm 8.0\times10^{-4}$ &  3.7  & $\rm
2.7\times 10^{-7}$ & $\rm
1.3\times 10^{38}$ \\
020331 & 35926 - 36835  & $\rm 4.7\times10^{-5}$ &  2.4  & $\rm
1.3\times 10^{-7}$ & $\rm
6.2\times 10^{37}$ \\
030212 & 64982 - 70775  & $\rm 1.8\times10^{-4}$ &  1.6  & $\rm
1.1\times 10^{-7}$ & $\rm
 5.3\times 10^{37}$ \\
\enddata
\end{deluxetable}


\clearpage

\begin{references}

\reference{}Barrio F.E., Done, C., \& Nayakshin, S., 2002, astro-ph/0209102

\reference{} Beloborodov, A.M., \& Illarionov, A.F., 2001 \mnras \, 323, 167

\reference{} Bolton, C. T., 1972, \nat \, 235, 271

\reference{} Borkus, V. et al., 1995, Astron. Lett. 21, 794

\reference{} Bowyer, S., Byram, E. T., Chubb, T., A., \& Friedman, M., 1965, Science 147, 394

\reference{} Brocksopp, C., Tarasov, A.E., Lyuty, V. M., and Roche, P., 1999, \aap \,343, 861

\reference{} Cui, W., Feng, Y-X., and Ertmer, M., 2002, \apj \, 564, L77

\reference{} Dove, J., Wilms, J., Nowak, M., Vaughan, B., and Begelman, M., 1998, \mnras \, 298, 729

\reference{} Esin, A., Narayan, R., Cui, W., Grove, J. E., and Zhang, S-N., 1998, \apj \, 505, 854

\reference{} Frontera, F., et al., 2001, \apj \, 546, 1027

\reference{} Giacconi, R., 1972, \apj \, 173, 79

\reference{} Gierlinski, M., et al., 1997, \mnras \, 288, 958

\reference{} Gierlinski, M., Zdziarski, A., Pountanen, J., Coppi, S., Ebisawa, K., and Johnson, W., 1999, \mnras \, 309, 958

\reference{} Gies, D., \& Bolton, C., 1986, \apj \, 304, 371 

\reference{} Golenetskii, S., Aptekar, R., Mazets, E., Frederiks, Cline, T., and Hurley, K., 2002a, \iaucirc\, 7940

\reference{} Golenetskii, S., Aptekar, R., Mazets, E., Frederiks, Cline, T., and Hurley, K., 2002b, GCN Circ. 1258
(http://gcn.gsfc.nasa.gov/gcn3/1258.gcn3)

\reference{} Hurley, K., Kouveliotou, C., Cline, T., Mazets, E., Golenetskii, S., Frederiks, D., and van Paradijs, J., 1999,
\apj \, 523, L37

\reference{} Illarionov, A.F., \& Sunyaev, R.A. 1975, \aap \, 39, 185

\reference{} Liang, E., and Nolan, P., 1984, Space Science Rev. 38, 353

\reference{} Ling, J., Mahoney, W., Wheaton, W., and Jacobson, A., 1987, \apj \, 321, L117

\reference{} Ling, J. et al., 1997, \apj \, 484, 375

\reference{}Maccarone, T., and Coppi, P., 2002, astro-ph/0204235

\reference{} Malzac, J., Beloborodov, A.M., \& Poutanen, J., 2001, \mnras \, 326, 417

\reference{} Mazets, E., 1995, in AIP Conf. Proc. 384, Gamma-Ray Bursts - 3rd Huntsville Symposium, Eds. C. Kouveliotou,
M. Briggs, \& G. J. Fishman (New York: AIP), 492

\reference{} Mason, P., McNamara, B., and Harrison, T., 1997, \aj \,114(1), 238 

\reference{}McConnell, M., et al., 2002, \apj \, 572, 984

\reference{} Petterson, J., 1978, \apj \, 224, 625 

\reference{} Phlips, B., et al., 1996, \apj \, 465, 907

\reference{}Poutanen, J., Krolik, J., and Ryde, F., 1997, \mnras \, 292, L21

\reference{}Romero, G., Kaufman Bernado, M., and Mirabel, I., 2002, \aap \, 393, L61

\reference{} Schmidt, M., 2002, \iaucirc\, 7856 

\reference{} Shakura, N. I. \& Sunyaev, R. A., 1973, \aap \, 24, 337 


\reference{} Stern, B., Poutanen, J., Svensson, R., Sikora, M., and Begelman, M.,
1995, \apj \, 449, L13

\reference{} Stern, B., Beloborodov, A., and Poutanen, J., 2001, \apj \, 555, 829

\reference{}Stirling, A., Spencer, R., de la Force, C., Garrett, M., Fender, R. and Ogley, R., 2001, \mnras \, 327, 1273

\reference{} Terekhov, M., Aptekar, R., Frederiks, D., Golenetskii, S., Il'inskii, V., and Mazets, E., 1998, in
AIP Conf. Proc. 428, Gamma-Ray Bursts - 4th Huntsville Symposium, Eds. C. Meegan, R., Preece, \& T. Koshut 
(New York: AIP), 894

\reference{} Webster, B. L., and Murdin, P., 1972, \nat \, 235, 37

\reference{} Zdziarski, A., Poutanen, J.,  Paciesas, W., and Wen, L., 2002, \apj \, 578, 357

\reference{} Zhang, S., Cui, W., Harmon, A., Paciesas, W., Remillard, R., and van Paradijs, J., 1997, \apj \,477, L95

\end{references}
\end{document}